\documentclass[twocolumn,showpacs,preprintnumbers,amsmath,amssymb,superscriptaddress]{revtex4}

\usepackage{graphicx}
\begin{document}
\bibliographystyle{prsty}
\title{Remote hole-doping of Mott insulators on the nanometer scale}

\author{M.~Takizawa}
\affiliation{Department of Physics, University of Tokyo, 3-7-1 Hongo, Bunkyo-ku, Tokyo, 113-0033, Japan}
\author{Y.~Hotta}
\affiliation{Department of Advanced Materials Science, University of Tokyo, 5-1-5 Kashiwanoha, Kashiwashi, Chiba, 277-8651, Japan}
\author{T.~Susaki}
\affiliation{Department of Advanced Materials Science, University of Tokyo, 5-1-5 Kashiwanoha, Kashiwashi, Chiba, 277-8651, Japan}
\author{Y.~Ishida}
\affiliation{Excitaion Order Research Team, RIKEN SPring-8 Center, Sayo-cho, Hyogo 679-5148, Japan}
\author{H.~Wadati}
\affiliation{Department of Physics, University of Tokyo, 3-7-1 Hongo, Bunkyo-ku, Tokyo, 113-0033, Japan}
\author{Y.~Takata}
\affiliation{Excitaion Order Research Team, RIKEN SPring-8 Center, Sayo-cho, Hyogo 679-5148, Japan}
\affiliation{Coherent X-ray Optics Laboratory, RIKEN SPring-8 Center, Sayo-cho, Hyogo 679-5148, Japan}
\author{K.~Horiba}
\affiliation{Excitaion Order Research Team, RIKEN SPring-8 Center, Sayo-cho, Hyogo 679-5148, Japan}
\author{M.~Matsunami}
\affiliation{Excitaion Order Research Team, RIKEN SPring-8 Center, Sayo-cho, Hyogo 679-5148, Japan}
\author{S.~Shin}
\affiliation{Excitaion Order Research Team, RIKEN SPring-8 Center, Sayo-cho, Hyogo 679-5148, Japan}
\author{M.~Yabashi}
\affiliation{Coherent X-ray Optics Laboratory, RIKEN SPring-8 Center, Sayo-cho, Hyogo 679-5148, Japan}
\affiliation{JASRI, SPring-8, Sayo-cho, Hyogo 679-5198, Japan}
\author{K.~Tamasaku}
\affiliation{Coherent X-ray Optics Laboratory, RIKEN SPring-8 Center, Sayo-cho, Hyogo 679-5148, Japan}
\author{N.~Nishino}
\affiliation{Coherent X-ray Optics Laboratory, RIKEN SPring-8 Center, Sayo-cho, Hyogo 679-5148, Japan}
\author{T.~Ishikawa}
\affiliation{Coherent X-ray Optics Laboratory, RIKEN SPring-8 Center, Sayo-cho, Hyogo 679-5148, Japan}
\affiliation{JASRI, SPring-8, Sayo-cho, Hyogo 679-5198, Japan}
\author{A.~Fujimori}
\affiliation{Department of Physics, University of Tokyo, 
3-7-1 Hongo, Bunkyo-ku, Tokyo, 113-0033, Japan}
\author{H.~Y.~Hwang}
\affiliation{Department of Advanced Materials Science, University of Tokyo, 5-1-5 Kashiwanoha, Kashiwashi, Chiba, 277-8651, Japan}
\affiliation{Japan Science and Technology Agency, Kawaguchi, 332-0012, Japan}

\date{\today}

\begin{abstract}
At interfaces between polar and nonpolar perovskite oxides, an unusual electron-doping has been previously observed, due to electronic reconstructions. 
We report on remote hole-doping at an interface composed of only polar layers, revealed by high-resolution hard x-ray core-level photoemission spectroscopy. 
In LaAlO$_3$/LaVO$_3$/LaAlO$_3$ trilayers, the vanadium valence systematically evolves from the bulk value of V$^{3+}$ to higher oxidation states with decreasing LaAlO$_3$ cap layer thickness. 
These results provide a synthetic approach to hole-doping transition metal oxide heterointerfaces without invoking a polar discontinuity. 
\end{abstract}


\maketitle
When assembled with atomic precision, the interface between polar and nonpolar perovskite oxides can be engineered to undergo new classes of interface electronic reconstructions, energetically driven by electrostatic boundary conditions \cite{HesperK3C60}. 
At such a polar discontinuity between two insulators, metallic, magnetic, and superconducting states have recently been discovered \cite{OhtomoLTOSTO,OkamotoLSTO,OhtomoLAOSTO,NakagawaLAOSTO,HuijbenLAOSTO,ThielLAOSTO,BrinkmanLAOSTO,Pentcheva-calc,Lee-calc,ReyrenLAOSTO}. 
For example, multilayers consisting of a band insulator SrTiO$_3$ and a Mott insulator LaTiO$_3$ exhibit metallic conductivity \cite{OhtomoLTOSTO,OkamotoLSTO,Pentcheva-calc,shibuya,TakizawaLTOSTO}. 
Interfaces between two band insulators, LaAlO$_3$ and SrTiO$_3$, also show metallic conductivity, and furthermore show a remarkable termination layer dependence \cite{OhtomoLAOSTO}. 
That is, the (LaO)$^+$/(TiO$_2$)$^0$ interfaces (``$n$-type'' interfaces) are metallic, 
while the (AlO$_2$)$^-$/(SrO)$^0$ interfaces (``$p$-type'' interfaces) remain insulating. 
In order to interpret such properties, ``electronic reconstructions'' to avoid the ``polar catastrophe'' have been proposed \cite{HesperK3C60,NakagawaLAOSTO,Lee-calc,polarReview}. 
For the (001) plane of LaAlO$_3$, where (LaO)$^+$ and (AlO$_2$)$^-$ are alternately stacked, in order to avoid the divergence of electrostatic potential with the number of layers, some charge redistribution must occur. 
For the (LaO)$^+$/(TiO$_2$)$^0$ interface, the divergence can be avoided if half an electron ($-$e/2) is added to the interfacial region through the change of the Ti valence from Ti$^{4+}$ to Ti$^{3.5+}$. 
Moreover, the metallic transport at the LaAlO$_3$/SrTiO$_3$ interfaces has been found to occur beyond a critical LaAlO$_3$ layer thickness of $\sim 4$ - 6 unit cells (uc) ($\sim 1.6$ - 2.3 nm) \cite{HuijbenLAOSTO,ThielLAOSTO}. 

This electronic reconstruction scenario should, in principle, be symmetric between electrons and holes. 
Despite significant effort, however, no examples of hole-doping have been experimentally demonstrated to date. 
The insulating $p$-type LaAlO$_3$/SrTiO$_3$ interface was found to be compensated by oxygen vacancies, rather than inducing holes \cite{NakagawaLAOSTO}. 
KTaO$_3$ grown on the TiO$_2$-terminated surface of SrTiO$_3$, also motivated to induce hole-doping, was found experimentally to have $n$-type carriers, likely originating from kinetically induced oxygen vacancies in the SrTiO$_3$ substrate \cite{KalabukhovKTO}. 
These negative results likely reflect the lack of energetically accessible higher oxidation states (such as Ti$^{5+}$) at these interfaces, and suggest that an interface must by suitably engineered to induce hole-doping. 
Furthermore, in the absence of demonstrable hole-doping, questions have been raised on the applicability of an electronic reconstruction picture as the origin of the electrons at the LaAlO$_3$/SrTiO$_3$ interface \cite{KalabukhovLAOSTO,SeimonsLAOSTO,HerranzLAOSTO}. 

Recently, LaAlO$_3$/LaVO$_3$ multilayers, which are composed of only polar planes, 
have been fabricated and studied by photoemission spectroscopy (PES) \cite{HottaLVO_PES,WadatiLVO}. 
It was found that the V $2p$ core-level spectra had not only V$^{3+}$ components as expected from the chemical composition, but also a higher oxidation state V$^{4+}$ component. 
Whether this V$^{4+}$ component originated simply from chemical imperfections or some electronic reconstruction mechanism was unclear. 
In this work, we have performed a systematic LaAlO$_3$ cap layer thickness dependence of the V valence in LaAlO$_3$/LaVO$_3$/LaAlO$_3$ trilayers using hard x-ray PES to probe deeply buried structures. 
The evolution of the vanadium valence with cap layer thickness demonstrates that heterointerfaces composed of only polar layers, free of any ionic polar discontinuity, can be remotely hole-doped depending on proximity to a polar surface. 

\begin{figure}
\begin{center}
\includegraphics[width=\linewidth]{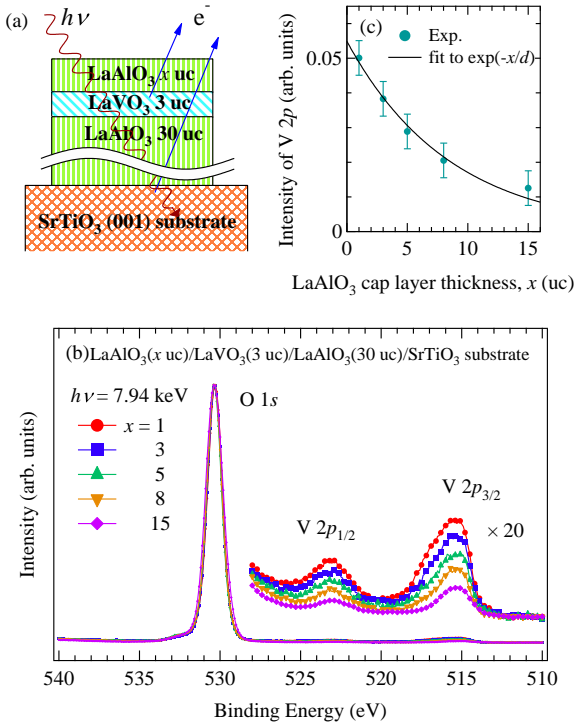}
\caption{(Color online) Schematic view of the trilayers and their core-level spectra. 
(a) Schematic view of the LaAlO$_3$/LaVO$_3$/LaAlO$_3$ trilayers grown on SrTiO$_3$ (001) substrates. 
We varied the LaAlO$_3$ cap layer thickness from one to 15 unit cells (uc) ($x = 1,\cdots,15$). 
(b) O~$1s$ and V~$2p$ core-level spectra of the LaAlO$_3$/LaVO$_3$/LaAlO$_3$ trilayers. 
(c) The intensity of the V~$2p$ core level as a function of LaAlO$_3$ cap layer thickness. 
The experimental data were well fit by the exponential function $a\exp(-x/d)$ with $d = 8.7 \pm 1$ uc. }
\label{sample}
\end{center}
\end{figure}
LaAlO$_3$($x$ uc)/LaVO$_3$(3 uc)/LaAlO$_3$(30 uc) trilayers, with varying LaAlO$_3$ cap layer thickness $x$, were grown on the atomically flat, TiO$_2$-terminated (001) surface of SrTiO$_3$ substrates using pulsed laser deposition (PLD), as schematically shown in Fig.~\ref{sample}~(a). 
All the trilayers were confirmed to be fully strained to the substrate by off-axis x-ray diffraction. 
The structures were grown at 600 $^{\circ}$C under an oxygen partial pressure of $1\times10^{−6}$ Torr, with a laser fluence of $\sim 2$ J/cm$^2$, following the previous optimization for two dimensional layer-by-layer growth of LaVO$_3$ \cite{HottaLVO}. 
Hard x-ray PES measurements were performed at the undulator beamline BL29XU of SPring-8, using a hemispherical electron energy analyzer, SCIENTA R4000-10kV. 
Details of the apparatus including x-ray optics are described elsewhere \cite{TakataHXPES,TamasakuHX,IshikawaHX}. 
Samples were transferred from the PLD chamber to the spectrometer chamber 
{\it ex situ} and no surface treatment was performed prior to PES measurements. 
All the measurements were carried out at room temperature, 
and the total energy resolution was set to about 200 meV. 
The Fermi level ($E_{\rm F}$) position was determined using gold spectra.

Figure~\ref{sample}~(b) shows the O~$1s$ and V~$2p$ core-level spectra of the LaAlO$_3$/LaVO$_3$/LaAlO$_3$ trilayers with varying LaAlO$_3$ overlayer thickness $x$, normalized to the O~$1s$ peak height. 
Because the LaVO$_3$ layer was only 3 uc thick, the V $2p$ core level signals were very small compared with those of O $1s$. 
Figure~\ref{sample}~(c) shows the integrated intensity of the V $2p$ core level as a function of LaAlO$_3$ cap layer thickness $x$. 
With increasing $x$, the intensity of the V $2p$ core level decreased, and the intensities were well fit by the exponential decay $a\exp(-x/d)$ with $d = 8.7 \pm 1$ uc ($\sim 3.5 \pm 0.4$ nm). 

\begin{figure}
\begin{center}
\includegraphics[width=\linewidth]{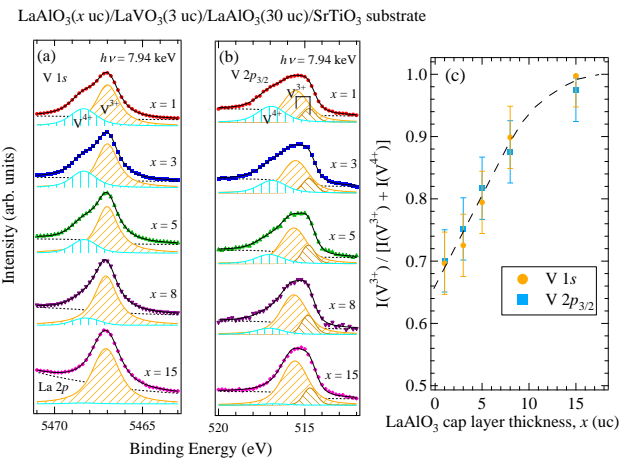}
\caption{(Color online) LaAlO$_3$ cap layer thickness dependence of the V core-level spectra of LaAlO$_3$($x$ uc)/LaVO$_3$(3 uc)/LaAlO$_3$(30 uc) trilayers grown on SrTiO$_3$ substrates. 
(a) V $1s$ core-level spectra and their line-shape analyses. 
(b) V $2p_{3/2}$ core-level spectra and their line-shape analyses. 
(c) The LaAlO$_3$ cap layer thickness dependence of the ratio of the V$^{3+}$ intensity to the total V intensity in the V $1s$ and $2p_{3/2}$ core-level spectra. 
The dashed curve is a guide to the eye. }
\label{Vcore}
\end{center}
\end{figure}

\begin{figure*}
\begin{minipage}{.65\linewidth}
\begin{center}
\includegraphics[width=\linewidth]{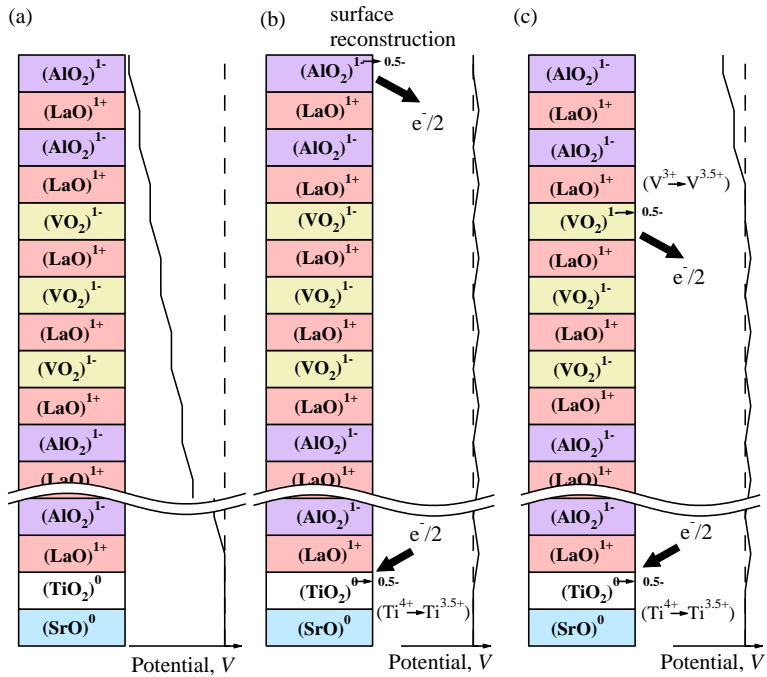}
\end{center}
\end{minipage}
\hfill
\begin{minipage}{.3\linewidth}
\caption{(Color online) Schematic illustrations of the reconstructions resolving the ``polar catastrophe'' in the all polar LaAlO$_3$/LaVO$_3$/LaAlO$_3$ trilayers. 
(a) The structures with all polar planes suffer from a ``polar catastrophe''. 
(b) The top surface of LaAlO$_3$ undergoes the usual atomic reconstruction, whereas the bottom LaAlO$_3$/SrTiO$_3$ interface accepts the extra electrons through the electronic reconstruction Ti$^{4+}$ $\rightarrow$ Ti$^{3.5+}$, thereby resolving the ``polar catastrophe''. 
(c) Alternatively, the valence change (V$^{3+}$ $\rightarrow$ V$^{3.5+}$) can also create extra electrons, resulting in purely electronic reconstructions on both sides of the trilayer. 
In the limit of a thick top LaAlO$_3$ layer, process (b) overcomes process (c). }
\label{polarcatastrophe}
\end{minipage}
\end{figure*}
Figure~\ref{Vcore}~(a) and (b) shows the V $1s$ and V $2p_{3/2}$ core-level spectra of the LaAlO$_3$/LaVO$_3$/LaAlO$_3$ trilayers and their line-shape decomposition. 
In both spectra, one can clearly see two components. 
The low and high binding energy components can be assigned to V$^{3+}$ and V$^{4+}$, respectively \cite{LCVO}. 
The V $1s$ and V $2p_{3/2}$ core-level spectra have been decomposed into two features by line-shape analysis. 
With increasing LaAlO$_3$ cap layer thickness, the structure on the higher binding-energy side (due to V$^{4+}$) decreases, meaning that as the LaVO$_3$ layer is more deeply buried in the LaAlO$_3$ environment, the V ion recovers the V$^{3+}$ character of bulk LaVO$_3$. 
In the line-shape analysis of the V $1s$ core-level spectra, an additional component from the La $2p$ core level located around $\sim 5485$ eV \cite{WadatiLVO} has also been taken into account [Fig.~\ref{Vcore}~(a)]. 
For the line-shape analysis of the V $2p_{3/2}$ core-level spectra, three components were needed to reproduce the experimental results well. 
One feature located around $\sim 517$ eV is the V$^{4+}$ component. 
Two other features located around $\sim 515$ eV and $\sim 516$ eV represent the multiplet structure of the V$^{3+}$ component [Fig.~\ref{Vcore}~(b)], as confirmed by cluster-model calculations including atomic multiplet structure. 
Figure~\ref{Vcore}~(c) shows the resulting V$^{3+}$ relative intensity as a function of LaAlO$_3$ cap layer thickness. 
With increasing thickness, the V$^{3+}$ component increases and saturates toward $\sim 1.0$ beyond $\sim 10$ uc. 

\begin{figure}[b]
\begin{center}
\includegraphics[width=.7\linewidth]{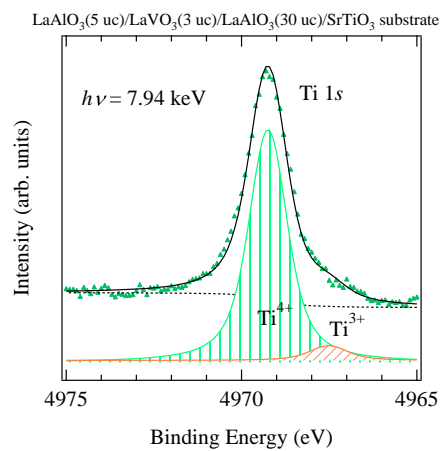}
\caption{(Color online) Typical Ti~$1s$ core-level spectrum of the LaAlO$_3$ (5 uc)/LaVO$_3$ (3 uc)/LaAlO$_3$ (30 uc) trilayer grown on an SrTiO$_3$ substrate. }
\label{Ticore}
\end{center}
\end{figure}

Let us discuss the origin of the present observations, together with the previous report that the valence distribution of V was highly asymmetric in the LaVO$_3$ layers in LaAlO$_3$/LaVO$_3$/LaAlO$_3$ \cite{HottaLVO_PES,WadatiLVO}. 
That is, V$^{4+}$ was preferentially distributed on the top side of the LaVO$_3$ layers. 
These features can be explained by considering the electrostatic potential of the trilayer, as schematically shown in Fig.~\ref{polarcatastrophe}. 
If all of the constituent materials preserve their bulk electronic and atomic configurations, the LaAlO$_3$/LaVO$_3$/LaAlO$_3$ trilayers films would consist of only polar planes and suffer from the ``polar catastrophe'' as shown in Fig.~\ref{polarcatastrophe}~(a). 
In order to avoid this, two types of reconstructions that dramatically alter the electrostatic potential may be possible. 
In the first scenario [Fig.~\ref{polarcatastrophe}~(b)], the polar (AlO$_2$)$^-$ surface of LaAlO$_3$ is reconstructed \cite{LAOsurf1,LAOsurf2}, resulting in the net ejection of the charge $-$e/2. 
Therefore, the LaVO$_3$ layer is not affected. 
In the second scenario [Fig.~\ref{polarcatastrophe}~(c)], the valence of the V ion changes from V$^{3+}$ to V$^{3.5+}$ and thereby $-$e/2 is removed from the top side of the embedded LaVO$_3$ layer. 
In either case, $-$e/2 is effectively transferred to the SrTiO$_3$ substrate. 
This can be seen in Fig.~\ref{Ticore}, showing typical Ti $1s$ core-level spectra from the film-substrate interface of the LaAlO$_3$/LaVO$_3$/LaAlO$_3$ trilayers. 
Despite the 38~uc thickness of the trilayer, Ti signals from the SrTiO$_3$ substrate were clearly observed due to the long electron escape depth using hard x-ray PES. 

Between these two scenarios, the energetically more favorable process will be realized for a given LaAlO$_3$ cap layer thickness. 
These two scenarios can be considered as two competing processes: 
In order to prevent the ``polar catastrophe'', some charge redistribution must occur, resulting in the removal of $-$e/2 from the top layer and transferred to the bottom of the polar mulilayers \cite{NakagawaLAOSTO}. 
When the LaAlO$_3$ cap layer thickness is thin, the electrostatic potential within the LaAlO$_3$ cap layer remains small and the valence change of V (V$^{3+}$ $\rightarrow$ V$^{4+}$) occurs [Fig.~\ref{polarcatastrophe}~(c)]. 
On the other hand, as the LaAlO$_3$ cap layers become thicker, the potential within the top LaAlO$_3$ cap layers grows in proportion to the LaAlO$_3$ cap layer thickness, and this increasing energy cost cannot be compensated simply by the valence change of V. 
Therefore, the reconstruction of the LaAlO$_3$ surface would suppress the potential divergence of the entire trilayer more effectively and the valence change of V becomes unnecessary – indeed, energetically unfavorable. 
Thus, the valence state of V in the LaVO$_3$ layers returns to its original bulk-like value of V$^{3+}$ [Fig.~\ref{polarcatastrophe}~(b)]. 

It is worth remarking on the relationship between the present observations and the thickness dependence of the interface carrier density in LaAlO$_3$/SrTiO$_3$ heterostructures, showing a critical LaAlO$_3$ layer thickness of 4 - 6 uc \cite{HuijbenLAOSTO,ThielLAOSTO}. 
This behavior may also be interpreted in the same scenario: 
For thin LaAlO$_3$ layer thickness, the electrostatic potential within the LaAlO$_3$ layer remains small, and therefore the redistribution of charge does not have to occur, resulting in few carriers induced at the LaAlO$_3$/SrTiO$_3$ interface. 
With increasing LaAlO$_3$ thickness, reconstruction of the LaAlO$_3$ surface donates $-$e/2 to the interfacial region, inducing carriers on the SrTiO$_3$ side. 
Whereas the LaAlO$_3$/SrTiO$_3$ structures induce electrons at the interface, by designing polar LaAlO$_3$/LaVO$_3$/LaAlO$_3$ trilayers, we have demonstrated remote hole-doping into a Mott insulator. 
This technique should be quite general, so long as electronic reconstructions are energetically favorable to surface reconstructions on short length scales. 
These results, together with previous studies of LaAlO$_3$/SrTiO$_3$ polar discontinuities, demonstrate the design of both positive and negative remote charge injection at heterointerfaces, opening a new door to the materials science and engineering of transition metal oxides. 

In conclusion, we have performed a hard x-ray photoemission spectroscopy study on LaAlO$_{3}$/LaVO$_{3}$/LaAlO$_{3}$ trilayers. 
The V core-level spectra showed two components, which we assigned to V$^{3+}$ and V$^{4+}$ valence states. 
The intensity of the V$^{3+}$ component increased with LaAlO$_{3}$ cap layer thickness and saturated beyond $\gg 10$ unit cells. 
This behavior can be explained by competing electronic reconstructions. 
This work demonstrates the ability to artificially hole-dope oxide heterostructures by atomic control of the electrostatic boundary conditions. 

We thank D.~A.~Muller and G.~A.~Sawatzky for fruitful discussions. 
This work was supported by a Grant-in-Aid for Scientific Research (A19204037) from the Japan Society for the Promotion of Science and a Grant-in-Aid for Scientific Research in Priority Areas ``Invention of Anomalous Quantum Materials'' from the Ministry of Education, Culture, Sports, Science and Technology. 
MT acknowledges support from the Japan Society for the Promotion of Science for Young Scientists.

\end{document}